\documentclass{iopart}

  \expandafter\let\csname equation*\endcsname\relax
  \expandafter\let\csname endequation*\endcsname\relax
\usepackage{amsmath,amssymb}
\usepackage{amsfonts}

\usepackage{bbm}
\usepackage{cite}
\usepackage{graphicx}
\usepackage[svgnames]{xcolor}
\usepackage[pdftex]{hyperref} 

\newcommand{\bvec}[1]{{\boldsymbol{#1}}}
\newcommand{\bk}{\bvec{k}}
\newcommand{\bn}{\bvec{n}}
\newcommand{\bp}{\bvec{p}}
\newcommand{\bq}{\bvec{q}}
\renewcommand{\br}{\bvec{r}}

\newcommand{\Z}{\mathbbm{Z}}

\newcommand{\avg}[1]{\overline{#1}}
\newcommand{\bea}{\begin{eqnarray}}
\newcommand{\eea}{\end{eqnarray}}

\newcommand{\dyadLambda}{\overleftrightarrow{\Lambda}}

\providecommand*{\I}{\mathrm{i}}                           
\providecommand*{\bra}[1]{\langle#1|}                      
\providecommand*{\ket}[1]{|#1\rangle}                      
\newcommand{\mIm}{\mathrm{Im}}				
\newcommand{\mRe}{\mathrm{Re}}				
\newcommand{\trace}{\mathrm{tr}}			

\renewcommand{\d}{\mathrm{d}}
\newcommand{\bok}[3]{\bra{#1} #2 \ket{#3}}

\begin{document}

\title[Semiclassical spectral function for waves in random potentials]{Semiclassical spectral function for matter waves in random potentials}

\author{M I Trappe$^1$, D Delande$^2$ and C A M\"uller$^{3,4}$}

\address{$^1$ Centre for Quantum Technologies, National University of
  Singapore, Singapore 117543, Singapore}
\address{$^2$ Laboratoire Kastler Brossel, UPMC-Paris 6, ENS, CNRS; F-75005 Paris, France}
\address{$^3$ Fachbereich Physik, Universit\"at Konstanz, 78457
  Konstanz, Germany}
\address{$^4$ INLN, Universit´e de Nice-Sophia Antipolis, CNRS; F-06560 Valbonne, France}

\eads{\mailto{cord.mueller@uni-konstanz.de }}

\begin{abstract}
An $\hbar$-expansion is presented for the ensemble-averaged spectral
function of noninteracting matter
waves in random potentials. We obtain the leading quantum
corrections to the deep classical limit at high energies by the Wigner-Weyl formalism. The analytical results are
checked with success against numerical data for Gaussian and laser speckle
potentials with Gaussian spatial correlation in two dimensions. 
\end{abstract}



\submitto{J. Phys. A: Math. Theor. }
\maketitle

\section{Introduction}

With this paper, we present a semiclassical calculation of the
spectral function that describes the joint energy-momentum
distribution of noninteracting (matter) waves in random potentials. This study has a
twofold motivation, theoretical as well as 
experimental. On the theoretical side, it is a 
challenge  in itself to obtain ensemble-averaged results by
semiclassical expansions in strongly disordered 
systems. The generic
case we study in detail is a quantum particle in a smooth, Gaussian random
process, a well-known model for
benchmarking different approximations 
\cite{Zimmermann2009}. 
On the experimental side, the dynamics of quantum particles
in random potentials is nowadays studied quite intensely with ultracold atoms in
random optical potentials, as reviewed in 
\cite{Modugno2010,Sanchez-Palencia2010,Shapiro2012}. 
Two issues under current scrutiny are the distinction
between classical diffusion and Anderson localisation
\cite{Kondov2011,McGehee2013,Mueller2014,McGehee2014}, 
and the determination of the mobility edge in laser speckle potentials \cite{Semeghini2014,Delande2014}. 
These potentials are a well-controlled
source of disorder with interesting statistical properties that 
differ from the simple Gaussian processes mentioned above
\cite{Kuhn2007,Lugan2009,Falco2010}. Practically, one has to know the spectral function precisely in order
to connect the experimentally observed momentum-space densities to
characteristic energies. 
In the strong-disorder regime, analytical weak-disorder
approximations cannot be applied.  Moreover, trustworthy numerical estimates
cost considerable computational resources, especially in
two- and three-dimensional settings. Thus, we
propose to study the spectral function in strong disorder potentials
by a systematic semiclassical expansion around the classical
solution.  

The rest of the article is structured as follows. Section
\ref{sec:spectralfctn} introduces the spectral function and its 
 properties relevant in the present context, together with the
ensemble-averaged density of states. We define the classical
limit and show why the usual weak-disorder approximations are
inadequate. In Section \ref{Semiclassicalcorrections} we provide the
computational framework for the leading quantum corrections, at least 
for the so-called Wigner-Weyl or smooth contribution of point-like
periodic orbits. The resulting formula is tested against numerical
data for the spectral function in a generic 2d Gaussian potential in
Section \ref{GaussPot}, and in a laser speckle potential in Section
\ref{SpecklePot}. We find excellent agreement for the Gaussian case, 
but also observe that the most salient quantum corrections for red- and especially blue-detuned speckle potentials are beyond the Wigner-Weyl approach.  Section \ref{sec:summary} summarises.

\section{Spectral function}
\label{sec:spectralfctn}

We define the spectral function for matter waves
in random potentials, describe a few of
its properties, and discuss the limitations of standard weak-disorder
approximations. 

\subsection{Definitions, properties}
We consider single-particle systems where the Hamiltonian generator of time evolution,
$H(\br,\bp) = T(\bp) + V(\br)$, is the sum of kinetic and potential
energy. In particular, let 
$V(\br)$ be a random potential for the real-space coordinate
vector $\br\in [0,L]^d =:L^d$ confined  to a $d$-dimensional cubic volume.  
$T(\bp)$ denotes the kinetic energy, a function of the
canonically conjugate momentum of the particle with mass
$m$. While our results pertain to 
arbitrary $T(\bp)$, we will use the Galilean
free-space dispersion $T(\bp) = \bp^2/2m$ in concrete examples, and
often use the index notation $T_\bp$. 

Heisenberg's commutation relation
imposes $[\hat\br,\hat\bp]=\I\hbar$ for the quantum mechanical observables.
Since potential and kinetic energy do not
commute, neither position nor momentum are good quantum numbers. One
possibility to describe the system is by a numerical diagonalisation
of the  random Hamiltonian for each realisation of disorder, followed
by an ensemble average, noted $\avg{(.)}$. However, this procedure can be very costly in terms of
computational resources, especially for high-dimensional, strongly
disordered systems that require many realisations to reach 
convergence. Analytical approaches rather try to describe ensemble-averaged quantities from the start. One such
quantity, the simplest in some sense, is the average single-particle resolvent
$\avg{\hat G}(z) = \avg{[z-\hat H]^{-1}}$, which is the object of the present
work.  

We will assume throughout that the disorder potential is
statistically homogeneous, meaning that the ensemble average
restores translation invariance. Consequently, momentum does become a
good quantum number for the average resolvent, whose matrix elements
define the single-particle
Green function $\bra{\bk}\avg{\hat G(z)}\ket{\bk'} = \delta_{\bk\bk'}
\avg{G}_{\bk}(z)$. By virtue of its analytical properties, the
Green function at any point $z$ in the complex energy plane, 
\bea
\avg{G}_\bk(z) = \int\d E \frac{A_\bk(E)}{z-E}, 
\eea
can be
reconstructed from its imaginary part on the real axis,
$A_\bk(E)=-\frac{1}{\pi}\mathrm{Im}\lim_{\epsilon\to0}\avg{G}_\bk(E+\I
  \epsilon)$.  
The latter function is known as the \textit{spectral function} and contains
vital information about the (averaged) spectrum of the
system. With help of the identity $\mIm(x-\I0)^{-1} = \pi \delta(x)$ for the
Dirac distribution, it can be written 
\begin{eqnarray}\label{AkE}
A_\bk(E)  = \bra{\bk}\avg{\delta(E - \hat H)}\ket{\bk}. 
\end{eqnarray}
It thus appears as the probability density that a plane-wave state
$\ket{\bk}$ has energy $E$. In the absence of a potential, the
spectral function $A_\bk(E) \stackrel{V=0}{\mapsto} \delta
(E-T_\bk)$ projects onto the free dispersion. As a rule, the stronger the
perturbation, the broader this function becomes. 

Summing over all states yields the
average density of states (AVDOS), 
\bea \label{avdos_def}
\nu(E) = \trace \hat A(E) = \sum_\bk A_\bk(E),
\eea
as the trace of the spectral
operator $ \hat A(E) = \avg{\delta(E-\hat H)}$. (Sums over momenta are
understood to run over $\bk = \hbar\bn/2\pi L$, $\bn\in\Z^d$ allowed by periodic
boundary conditions on $[0,L]^d$. Note that several conventions for the spectral function exist in the
literature, with factors of $2\pi$ in different places---see, e.g.,
\cite{Kuhn2007,Bruus2004} for an alternative.) The 
definition \eqref{AkE} implies the normalisation  
\bea\label{sumrule0}
\int\d E A_\bk(E) = 1. 
\eea
This relation is but the first out of a hierachy of \textit{sum
  rules} reading 
\bea\label{sumrulep}
a^p_\bk = \int \d E\, E^p A_\bk(E) = \bra{\bk}\avg{\hat H^p}\ket{\bk}. 
\eea
Apart from the normalisation $a_\bk^0=1$, equation \eqref{sumrule0}, we have the first moment $a_\bk^1 = T_\bk +
\avg{V(\br)}$. The constant 
\bea\label{meanV}
\avg{V(\br)}= :\avg{V}
\eea 
is the average value
of the random potential. Other moments of the on-site value are determined by the one-point
distribution function $P_1(V)$ according to  
\bea\label{P1}
\avg{V^m} =\int \d V \, V^m P_1(V). 
\eea  
Since we assume $V(\br)$ to be an ergodic process, ensemble averages can also be obtained
by spatial averages $\avg{V^m} =L^{-d}\int \d \br V(\br)^m$. 
(Integrals over positions are understood to run over the volume $[0,L]^d$.) 

With the second moment, 
\bea
a^2_\bk = \left(T_\bk+\avg{V}\right)^2 + \avg{ \delta V(\br)^2}. 
\eea
on-site fluctuations around the mean, $\delta
V(\br) = V(\br)-\avg{V}$, come into play. Their second moment, 
$ \avg{ \delta V(\br)^2}=:\delta V^2$, measures the on-site variance
of potential fluctuations. 

Starting from the third moment $a^3_\bk$, also 
information about spatial
correlations between fluctuations is encoded. In the remainder of this work, we will only require the  covariance
\bea\label{covarV}
\avg{\delta V(\br)\delta V(\br')} = \delta V^2C(\br-\br').
\eea   
Typically, the correlation function $C(\br)$  decays from 1 to 0 over a microscopic length
scale $\zeta$ (for notational simplicity,
we consider isotropic correlations). This introduces a correlation
energy scale, $E_\zeta = \hbar^2/m\zeta^2$. In the following, we will focus exclusively on the strong-potential regime defined by 
\bea\label{scregime}
\delta V\gg E_\zeta. 
\eea
In this situation, the atom kinetic energy fluctuates also by 
$\delta T(k) \sim \delta V \gg E_\zeta$, and therefore produces large
momenta $ k \gg
\hbar/\zeta$. In other words, the dynamics of matter waves inside the
random potential will be dominated by the \textit{semiclassical
  regime} $k \zeta \gg\hbar$. 

{In the following two sections, limiting cases are discussed where analytical results are known, namely the weak-disorder limit and the deep classical limit. }

\subsection{Self-energy and inadequacy of Born approximations}

{In the semiclassical regime, 
standard weak-disorder approximations  \cite{Kuhn2005,Kuhn2007,Yedjour2010,Piraud2013} prove to be inadequate.
The weak-disorder estimates try to} construct the spectral
function from the information contained in the lowest
moments of the random potential, mean \eqref{meanV} and covariance
\eqref{covarV}. This is most economically achieved by introducing the
self-energy $\Sigma_\bk(z)$ via Dyson's equation $\avg{G}_\bk(z) =
[z-T_\bk-\Sigma_\bk(z)]^{-1}$ and taking the limit $z=E+\I0$. 
The spectral function then is 
\bea\label{AfromSigma}
A_\bk(E)= -\frac{1}{\pi}
\frac{\mIm\Sigma_\bk(E)}{[E-{T_
\bk}-\mRe\Sigma_\bk(E)]^2+\mIm\Sigma_\bk(E)^2}.
\eea   
Formal operator identities permit to expand the
self-energy in an asymptotic series,  
\bea 
\Sigma_\bk(z) = \avg{V} + \bra{\bk}\left(\avg{\hat V (z-\hat T)^{-1}
    \hat V} -
  \avg{V}(z-\hat T)^{-1}\avg{V}\right) \ket{\bk} + \dots
\eea
involving only connected averages of the potential. The first term
merely shifts the energy by the potential mean. The
second term contains information about the variance. If
the series is truncated after this term, it results in the so-called
Born approximation (BA),  
\bea\label{SigmaBA}
\Sigma^\text{BA}_\bk(z) 
= \avg{V}+\sum_{\bk'} \frac{\avg{|\delta V_{\bk-\bk'}|^2}}{z-T_{\bk'}} 
= \avg{V}+\delta V^2 \sum_{\bk'} \frac{
  C_{\bk-\bk'}}{z-T_{\bk'}} 
\eea 
{where $\delta V_{\bk-\bk'} = \bra{\bk}\delta \hat V\ket{\bk'} = L^{-d} \int \rmd \br e^{-i(\bk-\bk')\cdot\br} \delta V(\br)$ is the Fourier component of the potential fluctuation. }
The second equality {of Eq.~\eqref{SigmaBA}} introduces the Fourier transform 
\bea\label{Ckcorr}
C_\bk=L^{-d}\int\d\br e^{-\I\bk\br}C(\br) =\avg{\delta V_\bk \delta V_\bk^*}/\delta V^2
\eea
 of the real-space correlator \eqref{covarV}, normalised to $\sum_\bk
 C_\bk=1$. 

Since the BA \eqref{SigmaBA} neglects terms of order $\delta V^3$, it can hold
only for weak potential fluctuations compared to the
other energies involved, $\delta V^2\ll T E_\zeta$ \cite{Kuhn2007}. 
A popular extension that pretends to 
a larger range of validity is the self-consistent BA
(SCBA), where the free propagator in \eqref{SigmaBA} is replaced by the
renormalised propagator itself, 
\bea\label{SigmaSCBA}
\Sigma^\text{SCBA}_\bk(z) 
= \avg{V}+\delta V^2 \sum_{\bk'} \frac{C_{\bk-\bk'}}
  {z-T_{\bk'}-\Sigma^\text{SCBA}_{\bk'}(z)}. 
\eea
This equation has to be solved numerically for the unknown
complex function $\Sigma^\text{SCBA}_\bk(z)$ appearing on both sides.   
Despite its popularity, the SCBA fails badly in the 
semiclassical regime of interest. Indeed, define 
$z_\bk:=(z-T_\bk-\avg{V})/\delta V$ as well as  $\sigma_\bk(z) :=(\Sigma_\bk(z)-\avg{V}) /\delta V$ and consider \eqref{SigmaSCBA} in the formal limit $\hbar\to0$
(i.e., $\delta V/E_\zeta\to \infty$). Since the correlation function
$C_{\bk-\bk'}$ constrains $|\bk-\bk'|$ to be of order $\hbar/\zeta$,
it turns into $\delta_{\bk\bk'}$. The scaled complex self-energy
$\sigma_\bk=\sigma'_\bk+\I\sigma''_\bk$ then  
obeys the self-consistent equation $\sigma_\bk= (z_\bk-\sigma_\bk)^{-1}$. This can be readily solved for the real and
imaginary parts, $\sigma'_\bk=z_\bk/2$ and $\sigma_\bk''= - 
[1-(z_\bk-\sigma_\bk')^2]^{1/2} = -\sqrt{1-z_\bk^2/4}$. As a
consequence, the spectral density \eqref{AfromSigma} becomes 
\bea
A_\bk^\text{SCBA}(E) = 
\frac{1}{\pi\delta V}\sqrt{1-\frac{(E-\avg{V}-T_\bk)^2}{4\delta
    V^2}}, \qquad (\hbar\to0), 
\label{semi-circle}
\eea
where $|E-\avg{V}-T_\bk|\le 2\delta V$, and vanishes elsewhere. 
In a zero-dimensional setting where the kinetic energy is
irrelevant, this form of the resulting AVDOS is known as  
``Wigner's semi-circle law,'' a celebrated property of random-matrix ensembles \cite{Wigner1955,Brody1981}. 
{However, this `universal' law generally lies far from the true result in the deep classical limit, discussed next.} 

\subsection{Classical limit}

One may neglect the non-commutativity of $\hat\br$ and $\hat\bp$
entirely in the deep classical limit $\delta V/E_\zeta\to\infty$ that
is noted commonly, if
rather abusively, $\hbar\to0$. In this limit, sometimes referred to as the
Thomas-Fermi limit \cite{Kane1963}, the expectation value 
\bea
\bra{\bk}\avg{\delta(E-\hat H)}\ket{\bk} \stackrel{\hbar\to0}{\to} 
\avg{\delta(E-T_\bk-V_1)}
\eea
 depends only on the potential value
$V_1=V(\br_1)$ at an arbitrary point $\br_1$. Then the ensemble average $\avg{\delta(x-V_1)} = \int
\d V_1 \delta(x-V_1) P_1(V_1) = P_1(x)$ produces the spectral function 
\bea\label{Acl} 
A_{\bk}^\text{cl}(E)=P_1(E-T_{\bk}). 
\eea 
The corresponding AVDOS \eqref{avdos_def} is 
\bea\label{avdoscl}
\nu^\text{cl}(E) = \sum_\bk P_1(E-T_\bk) = \int \rmd T
\nu_0(T) P_1(E-T),  
\eea
namely the convolution of the free DOS $\nu_0(E) $ with the one-point
distribution \cite{Kane1963,Falco2010}. For the Galilean dispersion, one has $\nu_0(E) = N_d
E^{(d-2)/2} \Theta(E)$, with $N_d$ a constant proportional to the $d$-dimensional volume of the system. 

The exact result \eqref{Acl} means that  
in the deep classical limit, the
probability for a particle with momentum $\bk$ to have potential energy
$V=E-T_\bk$ is given
by the on-site distribution $P_1(V)$ alone \cite{Pezze2011}. 
This distribution need not be close to the semi-circle law of the SCBA result, Eq.~\eqref{semi-circle}. 
Moreover, for certain classes of potentials such as the laser
speckle potentials discussed in section~\ref{SpecklePot} below, the distribution function is not even.  Then, there
can be no hope to catch the strong-disorder properties of the spectral function by an approximation like SCBA that is built solely upon even powers of the fluctuations, and a systematic 
description of quantum corrections beyond the classical limit is desirable. 
Quantum corrections to this classical limit arise because
finite  (in the sense of less-than-infinite) momenta probe nonlocal
features of the potential and thus should be sensitive to its
correlations. Accordingly, the classical limit \eqref{Acl} obeys the
first three sum rules for $p=0,1,2$ already on its own, and quantum corrections can
only be expected to arise for higher-oder sum rules $p\geq 3$, which are sensitive to
the spatial correlations. The
systematic semiclassical derivation of such corrections is the subject of the next 
section. 

\section{Semiclassical corrections}
\label{Semiclassicalcorrections}

With this section, we turn to the calculation of quantum
corrections 
\begin{eqnarray}
\Delta A_\bk(E) =A_\bk(E)- A^\text{cl}_\bk(E)
\end{eqnarray}
to the deep classical limit \eqref{Acl}. The proper tool
in the regime of interest, $\delta V\gg E_\zeta$, is a semiclassical
approximation. We face the task of computing the 
(momentum-)local average density of states, 
\begin{eqnarray}\label{AkEtrace}
A_{\bk}(E)=\mathrm{tr}\{\ket{\bk}\bra{\bk}\overline{\delta(E-\hat
  H)}\}.   
\end{eqnarray}
The calculation of traces like this has a long-standing history in
semiclassical physics, where phase-coherent quantum evolution is
described by the superposition of Feynman path amplitudes. Two types of contributions to
formal $\hbar$-expansions around the classical solution  are known
\cite{Gaspard1995,Richter2000}: first the
so-called smooth part, also known as Wigner-Weyl corrections, formally
due to orbits of zero length, and second
the so-called fluctuating part, due to periodic orbits of finite
length. 
{The present paper is devoted to the calculation of the smooth Wigner-Weyl corrections. 
The importance of periodic-orbit contributions is discussed in the concluding section \ref{sec:summary}. }

\subsection{Semiclassical expression for the spectral function}

In order to compute the quantum corrections to the smooth part, we employ Wigner's phase space formulation of quantum mechanics 
\cite{Wigner1932,Groenewold1946,Moyal1949,Imre1967,Wigner1984,Balazs1984,Berge1989}
in notations adapted to a finite-size system with discrete momenta
\cite{Gneiting2013,Fischer2013}. In this formalism, the trace of an
arbitrary operator $X(\hat{\br},\hat{\bp})$ is expressed as the
phase-space integral 
\begin{eqnarray}\label{trX}
\trace\{X(\hat{\br},\hat{\bp})\}=\frac{1}{L^d}\sum_{\bp}\int \d
\br\,X_W(\br,\bp) 
\end{eqnarray}
over its Wigner function 
\begin{eqnarray}\label{WignerA}
 X_W(\br,\bp) 
&=\int\d\br'\,
e^{\I \bp\br'/\hbar} \bok{\br-\frac{\br'}{2}}{X(\hat{\br},\hat{\bp})}{\br+\frac{\br'}{2}}
\\
\fl&=\sum_{\bq} e^{-2\I\bq\br/\hbar}\bok{\bp-\bq}{X(\hat{\br},\hat{\bp})}{\bp+\bq}.
\end{eqnarray}
The function $X_W(\br,\bp)$, also called Wigner transform
or Weyl symbol \cite{Zachos2005} can equivalently be written as the scalar
product  $X_W(\br,\bp) = \trace\{X(\hat{\br},\hat{\bp})W(\hat{\br}-\br;\hat\bp-\bp)\}$ 
with the Stratonovich-Weyl operator kernel
\begin{align}\label{WignerKernel}
W(\hat{\br}-\br;\hat\bp-\bp) & =\exp\left(-\frac{2\I}{\hbar}(\hat{\br}-\br);(\hat{\bp}-\bp)\right)\\
& = \int\d
\br'\,e^{\I\bp\br'/\hbar}\ket{\br+\frac{\br'}{2}}\bra{\br-\frac{\br'}{2}}\\
 & =\sum_{\bq}e^{-2\I\bq\br/\hbar}\ket{\bp+\bq}\bra{\bp-\bq}. 
\end{align}
The semicolon in (\ref{WignerKernel}) indicates ordering of products
of $\hat{\br}$ and $\hat{\bp}$ such that $\hat{\br}$ always stands
left of $\hat{\bp}$. Since this kernel obeys the Hilbert-Schmidt
orthogonality  
\begin{eqnarray}
\trace\{W(\hat{\br}-\br;\hat\bp-\bp)W(\hat{\br}-\br';\hat\bp-\bp')\}=\delta (\br-\br')\,L^d\,\delta_{\bp,\bp'}\ ,
\end{eqnarray}
one can invert \eqref{WignerA} and write the operator in terms of its Wigner
function,   
\begin{eqnarray}\label{OperAsWigner}
X(\hat{\br},\hat{\bp})=\frac{1}{L^d}\sum_{\bp}\int\d \br\,X_W(\br,\bp)\,W(\hat{\br}-\br;\hat\bp-\bp)\ .
\end{eqnarray}
Moreover, it follows that the trace of a product of operators $X(\hat{\br},\hat{\bp})$
and $Y(\hat{\br},\hat{\bp})$ is the phase-space integral of their
Wigner function product:  
\begin{eqnarray}\label{trAB2}
\trace\{\hat X\,\hat Y\}=\frac{1}{L^d}\sum_{\bp}\int\d \br\,X_W(\br,\bp)\,Y_W(\br,\bp)\ .
\end{eqnarray}
Applied to \eqref{AkEtrace} with $\hat X= \ket{\bk}\bra{\bk}$ and
  $\hat Y = \delta (E-\hat H)$ before the ensemble
  average, this yields 
\begin{eqnarray} 
\mathrm{tr}\{\ket{\bk}\bra{\bk}\delta(E-\hat H)\}
&=\frac{1}{L^d}\int\d \br\,[\delta(E-\hat H)]_W(\br,\bk) \label{trABeforeEnsemble}
\end{eqnarray}
since 
$\ket{\bk}\bra{\bk}_W(\br,\bp)=\delta_{\bk\bp}$ projects
onto the momentum $\bk$. 
Upon the ensemble average, the argument under the integral 
becomes independent of $\br$, and we thus arrive at the first result 
\begin{eqnarray}
A_\bk(E) 
= \avg{\delta(E-\hat H)_W}(\bk)\ .\label{trAafterEnsemble}
\end{eqnarray}

\subsection{Semiclassical expansion to order $\hbar^2$}

This previous result \eqref{trAafterEnsemble} is still exact. An approximation becomes necessary 
for the Wigner transform 
$\delta(E-\hat H)_W$, which cannot be determined in closed form for arbitrary
potentials. 
But one can compute quantum corrections to leading order
in $\hbar$, which are well known \cite{GrammaticosVoros1979}:
\begin{equation} 
\delta(E-\hat H)_W \approx \delta(E-H) 
-\frac{\hbar^2}{16} \left\{H  \overleftrightarrow{\Lambda}{}^2H \right\} \delta''(E-H) 
-\frac{\hbar^2}{24} \left\{H \overleftrightarrow{\Lambda} H
  \overleftrightarrow{\Lambda} H  \right\} \delta'''(E-H)
. 
\label{hbar2}
\end{equation} 
Here, $H = T(\bk)+V(\br)$ is the classical Hamiltonian function. The
first term inserted into \eqref{trAafterEnsemble} yields the
expected classical result \eqref{Acl}. Quantum corrections involve the differential operator 
\begin{eqnarray}
\dyadLambda=\overleftarrow{\partial_{\br}}\cdot\overrightarrow{\partial_{\bk}}-\overleftarrow{\partial_{\bk}}\cdot\overrightarrow{\partial_{\br}} \label{Lambda} 
\end{eqnarray}
that implements the Poisson bracket in linear order,
$\{f\overleftrightarrow{\Lambda}g\} = \{f,g\}$. 
In the second-order terms of \eqref{hbar2}, $\overleftrightarrow{\Lambda}$ is understood to
act only on the directly neighboring functions \cite{CinalBerge1993}. 
{The series  \eqref{hbar2} is obtained as a consequence of the Wigner representation of a product of operators, 
\begin{equation} 
[AB]_W(\br,\bp) = A_W(\br,\bp)\exp\left[\frac{i\hbar}{2}  \overleftrightarrow{\Lambda}\right]B_W(\br,\bp),  
\end{equation}
known as the Wigner-Groenewold-Moyal (star) product \cite{Wigner1932,Groenewold1946,Moyal1949}. 
}

For our separable Hamiltonian, we find 
\begin{eqnarray}
\fl  \big\{H \overleftrightarrow{\Lambda}{}^2H \big\}
&= 2\sum_{i,j=1}^d(\partial_{r_i}\partial_{r_j} V)(\partial_{k_{i}}\partial_{k_{j}}T)\ , \label{HLLH}\\
 \fl \big\{H\overleftrightarrow{\Lambda} H\overleftrightarrow{\Lambda} H\big\}
&=  - 
\sum_{i,j=1}^d 
\left[(\partial_{r_i}\partial_{r_j}
  V)(\partial_{k_{i}}T)(\partial_{k_{j}}T)+(\partial_{r_i}
  V)(\partial_{r_j} V)(\partial_{k_{i}}\partial_{k_{j}}T)\right]. \label{HLHLH}
\end{eqnarray}
We remark, moreover, that one may group the second terms
from \eqref{HLHLH}, of the type $(\partial V)^2\delta'''(E-H) = -
(\partial V) \partial \delta''(E-H)$, with 
the terms of type  $(\partial^2V) \delta''(E-H)$ from \eqref{HLLH} by  
partial integration under the $\br$-integral
in \eqref{trABeforeEnsemble}. 
Thus, the leading-order quantum
corrections to the spectral function are
\begin{eqnarray} 
\Delta A_{\bk}(E)\approx 
- \frac{\hbar^2}{12}\sum_{i,j=1}^d
\left[
m_{ij}^{-1}C_{ij}^{(2)}(\xi)
-\frac{v_i v_j}{2}C_{ij}^{(3)}(\xi)\right].
\label{AscR}
\end{eqnarray}
Here we have introduced the variable 
\begin{eqnarray}
\xi :=E-T_\bk
\end{eqnarray}
together with the possibly $\bk$-dependent tensors of inverse effective mass $m_{ij}^{-1} =
\partial_{k_i}\partial_{k_j}T_\bk$ and group velocities
$v_iv_j=(\partial_{k_i}T_\bk)(\partial_{k_j}T_\bk)$. 
The ensemble-averaged functions 
\begin{eqnarray}\label{Cijfuncxi}
C_{ij}^{(n)}(\xi):=\overline{(\partial_{r_i}\partial_{r_j} V(\br))\delta^{(n)}(\xi-V(\br))}
\end{eqnarray}
remain to be expressed by the statistical properties of the random process $V(\br)$.

\subsection{Moments from characteristic functional}

Expression \eqref{Cijfuncxi} requires to calculate moments of
potential derivatives. This suggests the Fourier
representation  
$V(\br)=\sum_{\bq}e^{\I\bq\br}V_{\bq}$
such that 
\begin{eqnarray}
\partial_{r_i}\partial_{r_j}V(\br)=-\sum_{\bq} q_i q_j e^{\I\bq\br}V_{\bq}. 
\end{eqnarray}
(We set $\hbar=1$ in the Fourier expansion since it does not interfere with the formal $\hbar$-expansion.) 
Similarly, we write the derivatives of Dirac distributions as the
Fourier integrals
\bea
\delta^{(n)} (\xi-V(\br)) = \partial_\xi^n\int\frac{\d\alpha}{2\pi} 
e^{-\I\alpha [\xi-V(\br)]}. 
\eea
Then, the ensemble average in \eqref{Cijfuncxi} has to be taken over the combination
$\overline{V_{\bq}\,e^{\I\alpha V(\br)}}$. 
We generate the prefactor $V_\bq$ by the differentiation  
\begin{eqnarray}
\overline{V_{\bq}\,e^{\I\alpha V(\br)}}
&=-\left.\I\frac{\partial\chi[\beta]}{\partial\beta_{\bq}}\right|_{\beta_{\bp}=\alpha
  e^{\I\bp\br}} \label{requiredFuncDiff}
\end{eqnarray}
of the characteristic functional in the momentum representation,
\begin{eqnarray}\label{genfunc}
\chi[\beta]=\overline{\exp \I\sum_{\bp}\beta_{\bp}V_{\bp}}. 
\end{eqnarray}
With the choice $\beta_\bp=\alpha e^{\I\bp\br}$, it returns  
\bea \label{chi1}
\chi[\alpha e^{\I\bp\br}] = \avg{\exp \I \alpha V(\br)}  = \int \d V
e^{\I\alpha V} P_1(V) =:
\chi_1(\alpha),   
\eea 
the characteristic function of the one-point
distribution. 
Since the ensemble-averaged result does not depend on the point $\br$,
we may choose $\br=0$ to simplify notations. 
Thus, the coefficients (\ref{Cijfuncxi}) become  
\begin{eqnarray}\label{C2E}
 C_{ij}^{(n)}(\xi)
&=\I\sum_{\bq}q_i q_j   \partial_\xi^n \int\frac{\d\alpha}{2\pi}e^{-\I\alpha\xi}
\left.\frac{\partial\chi[\beta]}{\partial\beta_{\bq}}\right|_{\beta_{\bp}=\alpha} \ .\label{C2Eb}
\end{eqnarray}
Here ends the general development of the theory. Once the
characteristic functional \eqref{genfunc} of a specific random
potential is given, one can compute the functions \eqref{C2E} and evaluate \eqref{AscR}. 
This will be carried out in the following two sections for a Gaussian
random process and a laser speckle potential, respectively. 

\section{Gaussian potentials}
\label{GaussPot}

As a benchmark test for the semiclassical approach we consider a
Gaussian random process centered on $\avg{V}=0$ and with variance
$V^2=\delta V^2$. 

\subsection{Statistical properties} 

The full distribution functional of the Gaussian random process of
interest is 
\begin{eqnarray}
P[V] 
&=\mathcal N_1\exp\left(-\frac{1}{2V^2}\sum_{\bq} V_{\bq}^* C_{\bq}^{-1}V_{\bq}\right)
\end{eqnarray}
with a normalisation constant $\mathcal N_1$ and the two-point
correlation \eqref{Ckcorr}. 
Its characteristic functional is equally Gaussian, 
\begin{eqnarray}
\chi[\beta]=\exp\left(-\frac{V^2}{2}\sum_{\bq}\beta_{\bq}^*\,C_{\bq}\beta_{\bq}\right)\ ,
\end{eqnarray}
and the derivative required in (\ref{C2Eb}) reads
\begin{eqnarray}\label{GaussDerivChi}
\left.\frac{\partial\chi[\beta]}{\partial\beta_{\bq}}\right|_{\beta_{\bp}=\alpha}=-V^2
C_{\bq} \alpha \chi_1(\alpha)
\end{eqnarray}
with $\chi_1(\alpha) = \exp(-V^2\alpha^2/2)$. 
The one-point distribution is of course also Gaussian, 
\bea \label{P1fromchi1}
P_1(\xi) = \int\frac{\d\alpha}{2\pi} e^{-\I\alpha\xi} \chi_1(\alpha) =
(2\pi V^2)^{-1/2}\exp \left(-\frac{\xi^2}{2V^2}\right). 
\eea

\subsection{Spectral function} 

With \eqref{GaussDerivChi} and \eqref{P1fromchi1}, the functions \eqref{C2E} become 
\begin{eqnarray}\label{C2EGauss}
C_{ij}^{(n)}(\xi) 
& = V^2 C_{ij}  P_1^{(n+1)}(\xi) 
\end{eqnarray}
where 
\bea \label{Cij}
C_{ij}=   \sum_\bq q_iq_j C_{\bq}
= - \partial_{r_i}\partial_{r_j}C(\br)|_{\br= 0} 
\eea
is the correlation curvature. 
Combining all factors, we obtain 
\bea
\Delta A_\bk(E) \approx - \frac{\hbar^2V^2}{12} \!\sum_{i,j=1}^d \!C_{ij}
  \left[m_{ij}^{-1} \partial_E^3 - \frac{v_iv_j}{2} \partial_E^4
  \right] \!P_1(E-T_\bk).  
\eea
With an isotropic dispersion $T_\bk=\bk^2/2m$ such that $m_{ij}^{-1}
= \delta_{ij}/m$ as well as $v_i=k_i/m$ and an 
isotropic Gaussian spatial correlation
$C(\br)=\exp\left[-\br^2/(2\zeta^2)\right]$ such that $C_{ij} =
\delta_{ij}/\zeta^2$, this
simplifies to 
\begin{eqnarray}\label{AGaussresult}
 \Delta A_{\bk}(E)\approx 
  -\frac{V^2E_\zeta}{12}\left[d\partial_E^3 -
    T_\bk\partial_E^4\right] P_1(E-T_\bk). 
\end{eqnarray}
This correction involves third and fourth derivatives of the one-point potential distribution, a property that ensures that quantum corrections do not change the 
sum rules \eqref{sumrulep} at the three lowest orders $p=0,1,2$, which are
alread exhausted by the classical limit $A^\text{cl}_\bk(E) = P_1(E-T_\bk)$. 
The first one to be corrected is the cubic moment, classically 
given by $a_\bk^{3,\text{cl}} = 3 V^2T_\bk+T_\bk^3$ and shifted  by $\Delta a_\bk^3 = dV^2E_\zeta/2$, actually independent of $\bk$,  
but governed by the ``quantum'' energy scale $E_\zeta=
\hbar^2/m\zeta^2$.

\begin{figure}
\begin{center}
\includegraphics[width=0.85\linewidth]{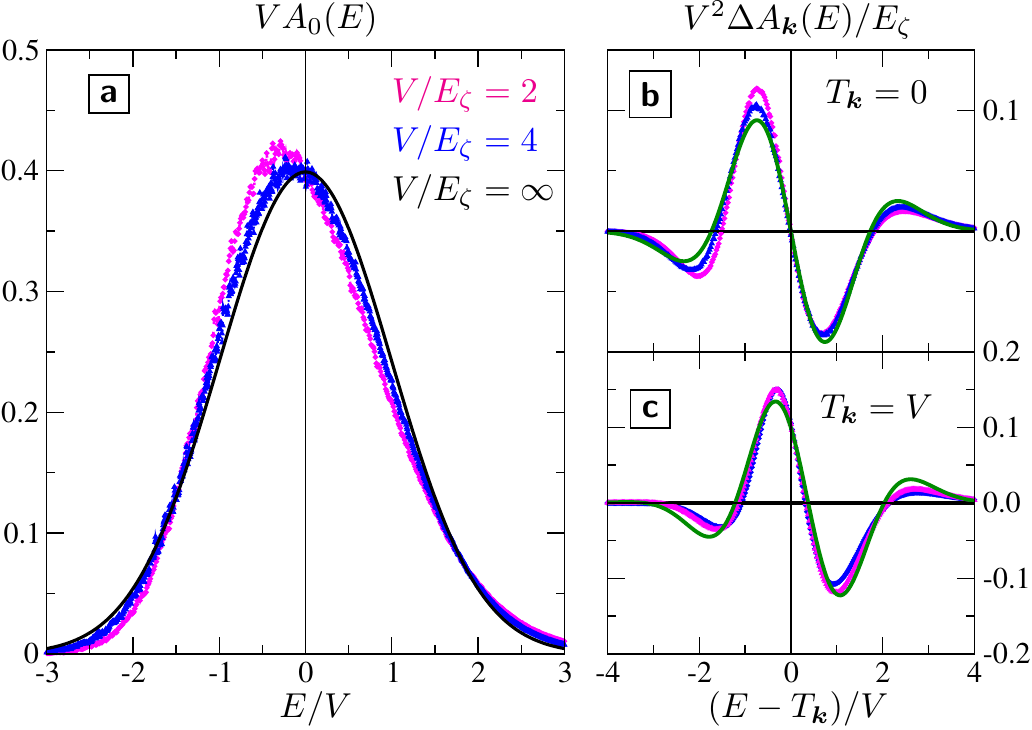}
\caption{
{Panel (a): Spectral function $A_0(E)$ at
zero momentum as function of energy $E$ (in units of rms potential
strength $V$) for $V/E_\zeta= 2,4$ in $d=2$ dimensions, approaching 
the normal distribution $P_1(E)$ (thick black line). 
Panel (b): The quantum correction $\Delta
A_0(E)$ in scaled vertical units such that the data collapse onto the
semiclassical approximation \eqref{A0EGauss}, shown as the continuous green 
curve on top of the data points. Panel (c): Same plot as in (b), but for finite momentum such that $T_\bk=V$, showing equally good agreement with the prediction \eqref{AGaussresult}.
}
}
\label{fig_A0E_gauss2d}
\end{center}
\end{figure}

Figure \ref{fig_A0E_gauss2d} shows in panel (a) $A_0(E)$ at
zero momentum as function of energy $E$ (both in units of the rms potential
strength $V$) for different values of $V/E_\zeta$ in $d=2$ dimensions. Data points are the
result of a numerical calculation and reproduce results available in
the literature \cite{Zimmermann2009}; \ref{sec:numerics} contains technical details about the numerical methods. 
The computed data curves converge toward the classical result, the normal distribution  
$A^\text{cl}_0(E) = P_1(E)$ of eq.~\eqref{P1fromchi1}. 
In panel (b), we plot the 
semiclassical
approximation for the quantum corrections, 
\bea\label{A0EGauss}
\Delta A_0(E) \approx - \frac{E E_\zeta}{6V^2}\left(3 - \frac{E^2}{V^2}\right)
P_1(E), 
\eea  
which is found to reproduce the data very well.  
Notably, since
the correction is proportional to $P_1'''(E)$, it vanishes at the origin 
and $E=\pm\sqrt{3}V$ and thus explains the approximate crossing of all curves at these points,
for large enough values of $V/E_\zeta$. The lowest-order approximation
starts to deviate from the data when $V/E_\zeta$
becomes too small, but the general trend is captured faithfully down to about 
$V/E_\zeta=2$.

{%
Indeed, since $P_1(\xi) = V^{-1}g(\xi/V)$ with a scalar gaussian function $g(x)$ of
order unity, 
 the correction \eqref{AGaussresult} scales as  
$E_\zeta/V^2$ multiplying another scalar function $h(x)$ of order unity (here a polynomial times $g(x)$). Our formal
$\hbar$-expansion is justified ex post if this correction is small,
which is the case in the semiclassical regime $E_\zeta/V\ll1$. Therefore, the quality of this approximation is independent of $T_\bk$, as apparent also from panel (c) in Fig.~\ref{fig_A0E_gauss2d}, where the deviation is plotted for finite $T_\bk=V$.   
}

\subsection{Average density of states}

In $d=2$ where $\nu_0(E) = N_2\Theta(E)$ is a pure  step function of energy, the classical AVDOS
\eqref{avdoscl} becomes the
integral of the one-point distribution up to $E$. For the normal
distribution, this gives the well-known error function, 
\bea
\nu^\text{cl}(E) = N_2 \int_{-\infty}^E \rmd T P_1(T)  =
\frac{N_2}{2}\left[1+\text{erf}(E/{\sqrt{2}}V)\right], 
\eea
which smoothens the free DOS around zero energy on the scale $V$. The
semiclassical corrections following from \eqref{AGaussresult} in this
case take a particularly simple form, 
\bea \label{delta_avdos}
\Delta\nu(E) \approx  -N_2\frac{V^2E_\zeta}{12} P_1''(E)  =
\frac{N_2E_\zeta}{{12}}\left({1}-\frac{E^2}{V^2}\right)P_1(E), 
\eea 
with a vanishing correction predicted at $|E|=V$. 


\section{Speckle potentials}
\label{SpecklePot}

Another class of interesting
random potentials are laser speckle, relevant for experiments with ultracold atoms \cite{Shapiro2012}. 
The atoms experience 
a dipolar light-shift potential
$V(\br)=s |E(\br)|^2$ proportional to the local light intensity
created by the random electric field $E(\br)$. The proportionality
factor $s$ encodes the atomic polarisability, besides the necessary
dimensionfull constants, and can have either positive or negative
sign, for blue- and red-detuned laser light, respectively. In the
following, we choose units such that $s=\pm 1$. We first review a few statistical properties and derive the
characteristic functional $\chi[\beta]$ needed for the
semiclassical corrections. 

\subsection{Statistical properties} 

We neglect finite-size and polarisation effects, and suppose that the
field $E(\br)$ is a scalar,
complex, Gaussian random process. Its real-space covariance is 
\bea
\avg{E^*(\br)E(\br')} =
\avg{|E|^2}\gamma(\br-\br') 
\eea  
with $\gamma(-\br) = \gamma(\br)^*$ and $\gamma(0)=1$ by definition. 
Equivalently,  the Fourier components  
$E_\bp = L^{-d} \int \d \br e^{-\I \bp\br}E(\br)$ and
$(E_{\bp})^*= (E^*)_{-\bp}=:E_{-\bp}^*$  are correlated by 
\bea
\avg{E^*_{-\bp} E_\bq} = \avg{|E|^2} \delta_{\bp\bq} \gamma_{-\bp}.
\eea
The momentum-space covariance components $\gamma_\bp=\gamma_\bp^*$ are normalised such that  
\bea\label{sumgammap1}
\sum_\bp \gamma_\bp= \gamma(0) = 1. 
\eea
The field distribution is
Gaussian, 
\bea
P[E,E^*] &= \mathcal N 
\exp\left\{ -\frac{1}{\avg{|E|^2}}\sum_{\bp} E^*_{-\bp}\gamma^{-1}_{-\bp} E_\bp
\right\} . 
\label{PEE}
\eea
Consequently, the potential
components $V_\bp = s\sum_\bq E^*_{-\bq} E_{\bq+\bp}$  have the characteristic functional 
\begin{eqnarray}
\fl \chi[\beta]& = \avg{\exp \I s \sum _{\bp,\bq} \beta_\bp E^*_{-\bq}
  E_{\bq+\bp}} 
= \int \mathrm{D}[E,E^*]
\exp\left(-\sum_{\bp,\bq}E^*_{-\bp}M_{\bp\bq}[\beta]E_{\bq}\right)\ . 
\end{eqnarray}
Here $D[E,E^*]$ is a suitably normalised measure, and 
the matrix $M_{\bp\bq}[\beta]$ is defined by 
\bea
M_{\bp\bq}[\beta] = \delta_{\bp\bq}\frac{\gamma_{-\bp}^{-1}}{\avg{|E|^2}} - \I s
  \beta_{\bq-\bp}
\eea 
Normalisation requires $\chi[0]\equiv 1$, and standard Gaussian
integration  results in 
\begin{eqnarray}
\chi[\beta]&=\frac{\det M[0]}{\det M[\beta]} = \exp[-\trace \ln (1+B[\beta])]\label{exp1+B}\ .
\end{eqnarray}
Here, the matrix  $B[\beta] = M[0]^{-1}M[\beta]-1$ has the elements
\bea
B[\beta]_{\bp\bq} = - \I \avg{V} \gamma_{-\bp}
\beta_{\bq-\bp}, 
\eea
where $s\avg{|E|^2} = \avg{V}$ has been used.


Before proceeding, let us obtain the on-site distribution by taking
$\beta_\bp=\alpha$ as in \eqref{chi1}. After expanding $\tr\ln(1+B)$
as a power series, we have to evaluate $\tr B^n$. But with the choice
$\beta_\bp=\alpha$ independent of $\bp$, this reduces to expressions
such as 
\bea
\tr B(\alpha)  = -\I \alpha \avg{V} \sum_\bp \gamma_\bp =  -\I\alpha  \avg{V}, 
\eea
and powers thereof, by virtue of the normalisation \eqref{sumgammap1}. 
As a result, we find 
\bea
\chi_1(\alpha) = \exp[-\ln (1-\I \alpha \avg{V})] = (1-\I\alpha \avg{V})^{-1}
\eea 
and from this deduce the one-point distribution \eqref{P1fromchi1} as   
\bea\label{P1speckle}
P_1(V) = \Theta(V/\avg{V})\frac{ e^{-V/\avg{V}}}{|\avg{V}|}. 
\eea
This is a one-sided exponential distribution on the positive real
axis for $\avg{V}>0$ and on the negative real axis for $\avg{V}<0$, as imposed
by the Heaviside distribution $\Theta(V/\avg{V})$. Its moments are
$\avg{V^m} = m! \avg{V}^m$, and thus the rms fluctuations are equal to the mean, $\delta V^2 = 2\avg{V}^2 - \avg{V}^2= \avg{V}^2$.

\subsection{Correlation moment}

We can now exploit the $\beta$-dependence of (\ref{exp1+B})
to calculate the derivative of $\chi[\beta]$, as needed for
(\ref{C2Eb}). 
As a first step we obtain
\begin{eqnarray}\label{SpeckleDerivChi1}
\frac{\partial}{\partial\beta_{\bq}}\chi[\beta]=g_{\bq}[\beta]\,\chi[\beta]\ ,
\end{eqnarray}
where
\begin{eqnarray}\label{gk}
g_{\bq}[\beta]= \sum_{l=1}^\infty(-1)^l 
\tr\left\{ \frac{\partial  B}{\partial\beta_\bq} B^{(l-1)}\right\} 
\end{eqnarray}
again after Taylor expansion of $\ln(1+B)$ and using the cyclic
property of the trace. The first term $l=1$ involves
\bea
\tr \frac{\partial  B}{\partial\beta_\bq} = -\I \avg{V}
\frac{\partial\beta_0}{\partial\beta_\bq} \sum_\bk\gamma_\bk  = -\I
\avg{V} \delta_{\bq0}
\eea 
and thus does not contribute to the $q_iq_j$-weighted sum in
\eqref{C2E}. Higher-order terms, when evaluated at $\beta_\bp=\alpha$ are found to be
\bea
\left.\tr\left\{ \frac{\partial  B}{\partial\beta_\bq}
    B^{(l-1)}\right\} \right|_\alpha = (-\I
\avg{V})^l\alpha^{l-1} \sum_\bk\gamma_\bk\gamma_{\bk+\bq}
\eea
where we recognise the potential correlator $C_\bq = \sum_\bk
\gamma_\bk\gamma_{\bk+\bq}$. The resulting correlation moment is 
\bea
\left.\frac{\partial\chi[\beta]}{\partial\beta_\bq}\right|_\alpha =
\frac{-\alpha \avg{V}^2}{(1-\I\alpha \avg{V})^2} C_\bq.
\eea

\subsection{Spectral function}

The functions $C_{ij}^{(n)}(\xi)$ of \eqref{C2E} then are
\bea
C_{ij}^{(n)}(\xi)& = -C_{ij} \partial_\xi^{n+1}
\int\frac{\d\alpha}{2\pi} \frac{e^{-\I\alpha \xi}}{(\alpha+\I/\avg{V})^2}
  \\
& =|\avg{V}| C_{ij}  \partial_\xi^{n+1} \left[\xi P_1(\xi)\right]
\eea
in terms of  \eqref{Cij} and \eqref{P1speckle}. 
These functions are to be contracted with the dispersion tensors according to
(\ref{AscR}). The semiclassical approximation for a Gaussian correlation with $C_{ij} =
\delta_{ij}/\zeta^2$ 
and dispersion
$T_\bk=\bk^2/2m$ therefore reads 
\begin{eqnarray}\label{Aresult}
\Delta A_\bk(E) \approx - \frac{E_\zeta |\avg{V}|}{12}\left[d\partial_E^3 -
  T_\bk\partial_E^4\right](E-T_\bk) P_1(E-T_\bk). 
\end{eqnarray}
Compared to the Gaussian potential case, \eqref{AGaussresult}, there is 
one factor $|\avg{V}|$ of rms potential strength less in front but one
more factor of $\xi=E-T_\bk$ under the derivatives;  we do not know whether this feature 
can be explained by a simple, intuitive argument.

\begin{figure}
\begin{center}
\includegraphics[width=0.75\linewidth]{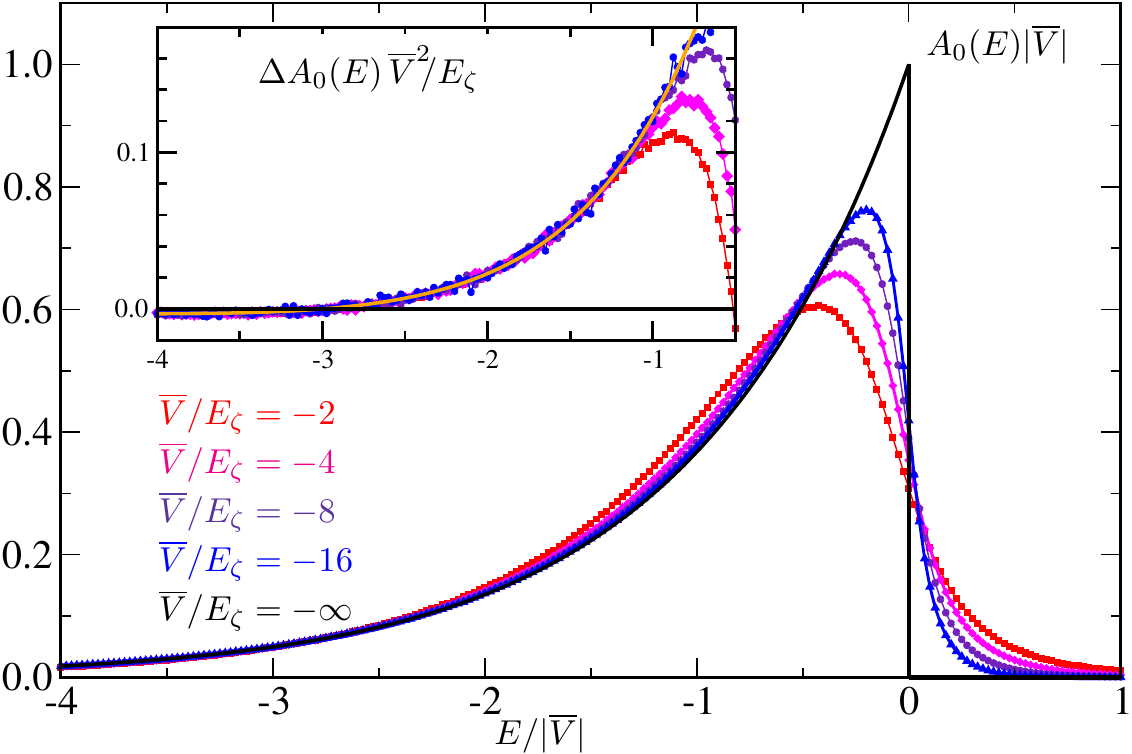}
\caption{Spectral function $A_0(E)$ at
zero momentum as function of energy $E$ (in units of rms potential
strength $|\avg{V}|$) in a red-detuned laser speckle potential
($\avg{V}<0$) with Gaussian
spatial correlation in $d=2$ dimensions. Main plot: The classical limit (thick black
line) strictly vanishes above $T_\bk$, here at zero energy. Quantum corrections round this discontinuity. 
Inset: The semiclassical
correction \eqref{DeltaA0Speckle} captures the smooth
behaviour away from the singularity;  the data points
for low enough energy collapse onto $(3-|x|)e^{-|x|}/6$, shown as the continous orange curve on top of the data points.  
} 
\label{fig_A0E_speckle2d_red}
\end{center}
\end{figure}

At zero momentum, the quantum correction reads 
\bea\label{DeltaA0Speckle}
\Delta A_0(E)  \approx   - \frac{d E_\zeta }{12 \avg{V}^2} f'''(E/\avg{V}) 
\eea
with $f(x) = \Theta(x) x e^{-x}$. Keeping in mind that $\delta(x)\varphi(x) \equiv \delta(x)\varphi(0)$, one 
finds $
f'''(x) = \delta'(x) - 2\delta(x) +\Theta(x) (3-x) e^{-x}
$ for the leading quantum corrections. 

\begin{figure}
\begin{center}
\includegraphics[width=0.75\linewidth]{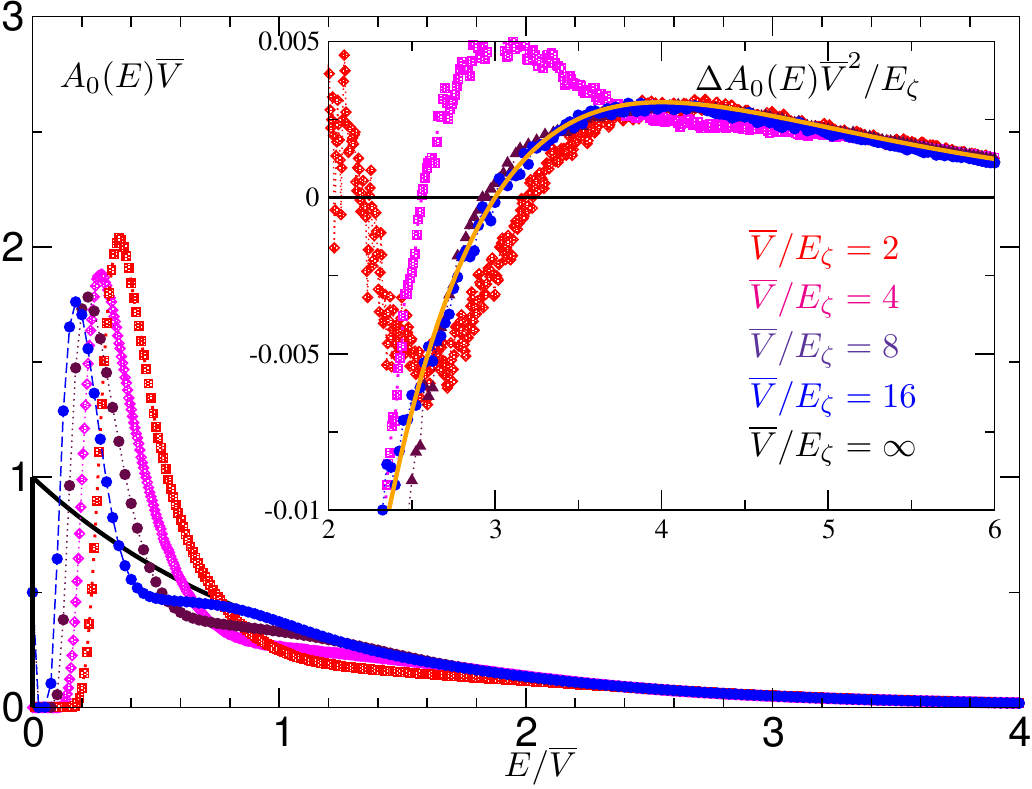}
\caption{Spectral function $A_0(E)$ at
zero momentum as function of energy $E$ (in units of rms potential
strength $\avg{V}$) in a blue-detuned laser speckle potential
($\avg{V}>0$) with Gaussian
spatial correlation in $d=2$ dimensions. Main plot: The classical limit (thick black
line) has a strict lower bound at zero energy, which induces large
singular quantum corrections. Inset: The semiclassical
correction \eqref{DeltaA0Speckle} captures the smooth
high-energy behaviour. Scales are chosen such that data points
progressively 
collapse onto $(x-3)e^{-x}$, shown as the continuous orange curve on top of the data points.  
} 
\label{fig_A0E_speckle2d_blue}
\end{center}
\end{figure}

Figure \ref{fig_A0E_speckle2d_red} shows numerical data for a
red-detuned speckle potential with $\avg{V}<0$ where the successive
curves approach the classical distribution, the rising exponential on
the negative real axis, with a discontinuity at the origin.   
The smooth quantum corrections  
away from the singular point are indeed captured correctly by
\eqref{DeltaA0Speckle}. Close to the origin, however, they are not
simply given by the singular terms $\propto 
[\delta'(x) - 2\delta(x)]$ of the Wigner-Weyl correction. Here, the quantum corrections are of a different nature. 
This behaviour is even more pronounced in a blue-detuned speckle
potential with $\avg{V}>0$, shown in
fig.~\ref{fig_A0E_speckle2d_blue}. The quantum corrections close to
the absolute lower bound of the classical distribution are highly
singular because they must preserve the support of the spectral
function on the positive real axis (since both kinetic and potential
energy are non-negative, contrary to the red-detuned case where
kinetic and potential energy can compensate each other).

\subsection{Average density of states}

The classical average density of states \eqref{avdoscl} in $d=2$ and for the
blue-detuned speckle is 
\bea
\nu^\text{cl}(E) = N_2 \Theta(E) (1 - e^{-E/\avg{V}}), \qquad
(\avg{V}>0).  
\eea
There are strictly no
states below zero because the potential is bounded from below, 
and the AVDOS then rises on the scale $\avg{V}$ to
its free value $N_2$. For the red-detuned speckle, one has 
\bea
\nu^\text{cl}(E) = N_2 \left[\Theta(-E) e^{-E/\avg{V}}+\Theta(E)\right], \qquad
(\avg{V}<0). 
\eea
Here, the rise to the bare DOS above zero energy happens for negative
energies. Both these behaviours are qualitatively very similar to the 1d case that
has been studied in detail by Falco et
al.~\cite{Falco2010}. Our systematic quantum correction to these
limits then is obtained by inserting \eqref{Aresult} into \eqref{avdos_def} and thus reads 
\bea
\Delta \nu(E) \approx -\frac{N_2 E_\zeta}{12|\avg{V}|} f''(E/\avg{V}),
\eea
with $f(x) = \Theta(x) xe^{-x}$ and thus $f''(x) =
\delta(x)+\Theta(x)(x-2)e^{-x}$. For the reasons explained in the
preceding section, we
cannot expect the singular correction around $E=0$ to be
accurate. However, at large enough distance from the singular point, the smooth correction
is proportional to $(E-2\avg{V})$ and thus
predicts an approximate crossing of curves at $E=2\avg{V}$.

\section{Summary and outlook}
\label{sec:summary}

In this paper, quantum corrections to the deep
classical limit of matter-wave spectral functions in random
potentials have been calculated using the Wigner-Weyl expansion for the smooth
contribution of point-like periodic orbits. These corrections are
expressed in closed form in terms of the one-point potential
distribution and its spatial covariance curvature. A comparison to numerical data for two-dimensional systems
reveals that the leading-order Wigner-Weyl corrections apply, as expected, with
quantitative agreement to generic Gaussian-distributed potentials. But
we also observe that for laser speckle potentials, the
smooth corrections only apply in the large-energy sector, where the spectral weight is rather small. 

{%
So-called oscillatory contributions from 
periodic orbits of finite length have been neglected throughout. 
This approximation can be
expected to be valid whenever these orbits depend
very sensitively upon the detailed potential configuration for each
realisation of disorder. An average over either an energy range or an
ensemble of realisations then wipes out these
fluctuations. 
Obviously, this approximation works well with the Gaussian
random potential of section~\ref{GaussPot}, where the main spectral
weight lies at energies $|E|\lesssim \delta V$ around the mean potential, much above the deep wells. 
In contrast, for the speckle potential of section~\ref{SpecklePot} 
quantum corrections beyond the smooth Wigner-Weyl terms are important, because a rather large spectral weight is located close to $E=0$, where the classical distribution is discontinuous, and corrections are of a more singular nature.  
Indeed, their physical origin then is the zero-point 
energy shift $\hbar\omega/2$ of locally bound states, where $\omega \propto \sqrt{V''_0}$ is of the order of the local harmonic
oscillator frequency.
In principle, such a correction can be described as the result of finite-size periodic orbits, located in the local potential minima. 
These short trajectories at low energies are all very similar to each other, 
and their effect can survive the ensemble average. 
The quantitative treatment of these corrections is
beyond the scope of this paper and remains a subject for future research.  
 }

\ack

Enlightening discussions with B Gr\'emaud, C Miniatura, J-D Urbina,
and K Richter are gratefully acknowledged. 
C.A.M. acknowledges the hospitality of Universit\'e 
Pierre et Marie Curie and Laboratoire Kastler Brossel, 
Paris. This work was granted access to the HPC resources of TGCC under the allocation 2014-056089 made by GENCI 
(Grand Equipement National de Calcul Intensif) and to the HPC resources of The Institute 
for Scientific Computing and Simulation financed by Region Ile de France and the project Equip@Meso (reference ANR-10-EQPX- 29-01).

\appendix

\section{Numerical methods} 
\label{sec:numerics} 

In order to test our analytical predictions, we have performed extensive numerical calculations
of the spectral function for various types of disorder in $d=2$ dimensions.
For the Gaussian-distributed potential discussed in section \ref{GaussPot}, our data reproduce results available in the literature~\cite{Zimmermann2009}.
The starting point is \eqref{AkE} and the following
temporal representation of the $\delta$ function:
\begin{equation}
\delta(E-\hat{H}) = \frac{1}{\pi} \mIm \frac{1}{E-\hat{H}-\rmi 0} = \frac{1}{\pi} \mRe \int_0^{\infty}{\rme^{-\rmi \hat{H}t}\ \rme^{\rmi Et}\ \rmd t}, 
\end{equation}
which give
\begin{equation}
 A_{\bk}(E) = \frac{1}{\pi} \mRe \int_0^{\infty}{\avg{\langle \bk|\rme^{-\rmi \hat{H}t}|\bk \rangle} \ \rme^{\rmi Et}\ \rmd t}.
\end{equation}
The numerical calculation then amounts to propagating an initial plane wave $|\bk\rangle$ with the disordered Hamiltonian $\hat{H}$ during time $t$ and to computing
the overlap of the time evolved state with $\langle \bk|,$ followed by a Fourier transform from time to energy. This step must be repeated for
several independent realizations of the disorder in order to perform disorder averaging.
In order to obtain small statistical fluctuations even in the tail of the spectral function, a rather large number (more than 120 000) of disorder realizations
was used. 
An alternative numerical method~\cite{Semeghini2014} consists in computing eigenstates close to $E$ and determining their momentum representation, followed by an ensemble average. The latter method turns out to consume much more computational resources.

The system is first discretized on a 2D grid of size $L\times L$ with periodic boundary conditions along $x$ and $y$. The spatial discretization step must be much smaller than both the correlation length of the disordered potential and the typical de Broglie 
wavelength of the propagated state. Typically, a cell of surface $(\pi\zeta)^2$ is discretized in 8-20 steps (depending on the disorder strength) along both $x$ and $y$. 
A spatially correlated complex Gaussian field is generated on the
grid in a standard way, by convoluting a spatially uncorrelated complex Gaussian
field with a proper cutoff function~\cite{KuhnPhD2007,Delande2014}. 
The Gaussian correlated potential is obtained by simply taking the
real part of the complex field. The speckle potential is obtained by
taking the squared modulus of the complex field.

Also, the system size must be chosen much larger than the scattering mean free path at all energies where the spectral function is significant. In practice,
a size $L=20\pi\zeta$ was found sufficient.

The CPU-consuming part of our calculation is the temporal propagation of the initial state $|\bk\rangle$ with the disordered Hamiltonian $\hat{H}.$ The time propagation itself uses an iterative method based on the expansion of the evolution operator in combinations of Chebyshev polynomials of the 
Hamiltonian \cite{Roche1997,Fehske2009}. This procedure is repeated for many disorder realizations, which finally gives access to the 
spectral function. Since the spectral function is a smooth function of energy, its Fourier transform 
decays relatively fast at long times.
This makes it possible to numerically propagate the initial state $|\bk\rangle$ only over a restricted time interval, which reduces the computing time substantially.



\section*{References}

\bibliographystyle{h-physrev} 
\bibliography{semiclAkE}

\begin{thebibliography}{10}

\bibitem{Zimmermann2009}
R.~Zimmermann and C.~Schindler,
\newblock Phys. Rev. B {\bf 80}, 144202 (2009).

\bibitem{Modugno2010}
G.~Modugno,
\newblock Rep. Progr. Phys. {\bf 73}, 102401 (2010).

\bibitem{Sanchez-Palencia2010}
L.~Sanchez-Palencia and M.~Lewenstein,
\newblock Nat. Phys. {\bf 6}, 87 (2010).

\bibitem{Shapiro2012}
B.~Shapiro,
\newblock J. Phys. A: Math. Theor. {\bf 45}, 143001 (2012).

\bibitem{Kondov2011}
S.~S. Kondov, W.~R. McGehee, J.~J. Zirbel, and B.~DeMarco,
\newblock Science {\bf 334}, 66 (2011).

\bibitem{McGehee2013}
W.~R. McGehee, S.~S. Kondov, W.~Xu, J.~J. Zirbel, and B.~DeMarco,
\newblock Phys. Rev. Lett. {\bf 111}, 145303 (2013).

\bibitem{Mueller2014}
C.~A. M\"{u}ller and B.~Shapiro,
\newblock Phys. Rev. Lett. {\bf 113}, 099601 (2014).

\bibitem{McGehee2014}
W.~R. McGehee, S.~S. Kondov, W.~Xu, J.~J. Zirbel, and B.~DeMarco,
\newblock Phys. Rev. Lett. {\bf 113}, 099602 (2014).

\bibitem{Semeghini2014}
G.~Semeghini {\em et~al.},
\newblock {Measurement of the mobility edge for 3D {Anderson} localization},
  2014, arXiv:1404.3528.

\bibitem{Delande2014}
D.~Delande and G.~Orso,
\newblock Phys. Rev. Lett. {\bf 113}, 060601 (2014).

\bibitem{Kuhn2007}
R.~Kuhn, O.~Sigwarth, C.~Miniatura, D.~Delande, and C.~A. M\"{u}ller,
\newblock New J. Phys. {\bf 9}, 161 (2007).

\bibitem{Lugan2009}
P.~Lugan {\em et~al.},
\newblock Phys. Rev. A {\bf 80}, 023605 (2009), arXiv:0902.0107.

\bibitem{Falco2010}
G.~M. Falco, A.~A. Fedorenko, J.~Giacomelli, and M.~Modugno,
\newblock Phys. Rev. A {\bf 82}, 053405 (2010).

\bibitem{Bruus2004}
H.~Bruus and K.~Flensberg,
\newblock {\em Many-Body Quantum Theory in Condensed Matter Physics} (Oxford
  Univ. Press, 2004).

\bibitem{Kuhn2005}
R.~C. Kuhn, C.~Miniatura, D.~Delande, O.~Sigwarth, and C.~A. M\"uller,
\newblock Phys. Rev. Lett. {\bf 95}, 250403 (2005).

\bibitem{Yedjour2010}
A.~Yedjour and B.~A. Tiggelen,
\newblock Eur. Phys. J. D {\bf 59}, 249 (2010).

\bibitem{Piraud2013}
M.~Piraud, L.~Pezz\'e, and L.~Sanchez-Palencia,
\newblock New J. Phys. {\bf 15}, 075007 (2013).

\bibitem{Wigner1955}
E.~P. Wigner,
\newblock Ann. Mathem. {\bf 62}, pp. 548 (1955).

\bibitem{Brody1981}
T.~A. Brody {\em et~al.},
\newblock Rev. Mod. Phys. {\bf 53}, 385 (1981).

\bibitem{Kane1963}
E.~O. Kane,
\newblock Phys. Rev. {\bf 131}, 79 (1963).

\bibitem{Pezze2011}
L.~Pezz\'e {\em et~al.},
\newblock New J. Phys. {\bf 13}, 095015 (2011).

\bibitem{Gaspard1995}
P.~Gaspard,
\newblock $\hbar$-expansion for quantum trace formula,
\newblock in {\em Quantum Chaos: Between Order and Disorder}, edited by
  G.~Casati and B.~Chirikov, pp. 385--404, Cambridge Univ. Press, 1995.

\bibitem{Richter2000}
K.~Richter,
\newblock {\em Semiclassical Theory of Mesoscopic Quantum Systems} (Springer
  Berlin Heidelberg, 2000).

\bibitem{Wigner1932}
E.~Wigner,
\newblock Phys. Rev. {\bf 40}, 749 (1932).

\bibitem{Groenewold1946}
H.~J. Groenewold,
\newblock Physica {\bf 12}, 405 (1946).

\bibitem{Moyal1949}
J.~E. Moyal,
\newblock Math. Proc. Camb. Phil. Soc. {\bf 45}, 99 (1949).

\bibitem{Imre1967}
K.~Imre, E.~\"{O}zizmir, M.~Rosenbaum, and P.~F. Zweifel,
\newblock J. Math. Phys. {\bf 8}, 1097 (1967).

\bibitem{Wigner1984}
M.~Hillary, R.~F. O'Connell, M.~O. Scully, and E.~P. Wigner,
\newblock Phys. Rep. {\bf 106}, 121 (1984).

\bibitem{Balazs1984}
N.~L. Balazs and B.~K. Jennings,
\newblock Phys. Rep. {\bf 104}, 347 (1984).

\bibitem{Berge1989}
B.-G. Englert,
\newblock J. Phys. A: Math. Gen. {\bf 22}, 625 (1989).

\bibitem{Gneiting2013}
C.~Gneiting, T.~Fischer, and K.~Hornberger,
\newblock Phys. Rev. A {\bf 88}, 062117 (2013).

\bibitem{Fischer2013}
T.~Fischer, C.~Gneiting, and K.~Hornberger,
\newblock New J. Phys. {\bf 15}, 063004 (2013).

\bibitem{Zachos2005}
C.~K. Zachos, D.~B. Fairlie, and T.~L. Curtright,
\newblock {\em Quantum Mechanics in Phase Space} (World Scientific, 2005).

\bibitem{GrammaticosVoros1979}
B.~Grammaticos and A.~Voros,
\newblock Ann. Phys. {\bf 123}, 359 (1979).

\bibitem{CinalBerge1993}
M.~Cinal and B.-G. Englert,
\newblock Phys. Rev. A {\bf 48}, 1893 (1993).

\bibitem{KuhnPhD2007}
R.~Kuhn,
\newblock {\em Coherent Transport of Matter Waves in Disordered Optical
  Potentials},
\newblock PhD thesis, Universit\"at Bayreuth, 2007.

\bibitem{Roche1997}
S.~Roche and D.~Mayou,
\newblock Phys. Rev. Lett. {\bf 79}, 2518 (1997).

\bibitem{Fehske2009}
H.~Fehske {\em et~al.},
\newblock Phys. Lett. A {\bf 373}, 2182  (2009).

\end{thebibliography}

\end{document}